\documentclass[aps,prl,twocolumn,superscriptaddress,showpacs]{revtex4-1}

\usepackage{amsmath,amssymb,graphics,epsfig,epstopdf,color,verbatim,ulem,braket,tabularx}
\usepackage{multirow}
\usepackage[colorlinks,linkcolor=blue,citecolor=blue,urlcolor=blue,bookmarks=false]{hyperref}

\usepackage{listings}

\begin{document}

\title{Higher-order Topology of Axion Insulator EuIn$_2$As$_2$}

\author{Yuanfeng Xu}
\affiliation{Beijing National Laboratory for Condensed Matter Physics, and Institute of Physics, Chinese Academy of Sciences, Beijing 100190, China}
\affiliation{University of Chinese Academy of Sciences, Beijing 100049, China}
\affiliation{Max Planck Institute of Microstructure Physics, 06120 Halle, Germany}

\author{Zhida Song}
\affiliation{Beijing National Laboratory for Condensed Matter Physics, and Institute of Physics, Chinese Academy of Sciences, Beijing 100190, China}
\affiliation{University of Chinese Academy of Sciences, Beijing 100049, China}
\affiliation{Department of Physics, Princeton University, Princeton, New Jersey 08544, USA}

\author{Zhijun Wang}
\affiliation{Beijing National Laboratory for Condensed Matter Physics, and Institute of Physics, Chinese Academy of Sciences, Beijing 100190, China}
\affiliation{University of Chinese Academy of Sciences, Beijing 100049, China}

\author{Hongming Weng}
\email{hmweng@iphy.ac.cn}
\affiliation{Beijing National Laboratory for Condensed Matter Physics, and Institute of Physics, Chinese Academy of Sciences, Beijing 100190, China}
\affiliation{University of Chinese Academy of Sciences, Beijing 100049, China}
\affiliation{CAS Center for Excellence in Topological Quantum Computation, University of Chinese Academy of Sciences, Beijing 100190, China}
\affiliation{Songshan Lake Materials Laboratory, Dongguan, Guangdong 523808, China}

\author{Xi Dai}
\email{daix@ust.hk}
\affiliation{Department of Physics, Hong Kong University of Science and Technology, Clear Water Bay, Kowloon, Hong Kong}

\begin{abstract}
Based on first-principles calculations and symmetry analysis, 
we propose that EuIn$_2$As$_2$ is a long awaited axion insulator with antiferromagnetic (AFM) long range order. Characterized by the parity-based invariant $\mathbb Z_4=2$, the topological magneto-electric effect is quantized with $\theta=\pi$ in the bulk, with a band gap as large as 0.1 eV. 
When the staggered magnetic moment of the AFM phase is along $a/b$-axis, it's also a TCI phase. Gapless surface states emerge on (100), (010) and (001) surfaces, protected by mirror symmetries ({\it nonzero} mirror Chern numbers). When the magnetic moment is along $c$-axis, the (100) and (001) surfaces are gapped. 
As a consequence of a high-order topological insulator with $\mathbb Z_4=2$, the one-dimensional (1D) chiral state can exist on the hinge between those gapped surfaces. We have calculated both the topological surface states and hinge state in different phases of the system, respectively, which can be detected by ARPES or STM experiments. 
\end{abstract}

\maketitle

\textit{Introduction. }The concept of axion field is introduced to solve the strong charge-parity (CP) problem in quantum chromodynamics. ~\cite{peccei1977cp}
While in condensed matter physics, axion field appears in the field theory description of the topological magneto-electric(TME) effect,~\cite{wilczek1987two,qi2008topological} with an effective action in the form,
\begin{equation}
S_{\theta}=\frac{\theta e^2}{4\pi^2}\int dt d^{3}x \textit{\textbf{E}} \cdot \textit{\textbf{B}}
\end{equation}
where $\textit{\textbf{E}}$ and $\textit{\textbf{B}}$ are electromagnetic fields and the coefficient $\theta$ is axion angle with a period of $2\pi$. 
This term will modify the Maxwell's equation in classical electrodynamics and lead to the TME effect.~\cite{wilczek1987two} 
As the magnetoelectric coupling term $\textit{\textbf{E}}\cdot \textit{\textbf{B}}$ changes sign under time reversal symmetry(TRS) $\mathcal{T}$ or 
inversion symmetry $\mathcal{I}$, the only allowed value of $\theta$ has to be quantized to 0 or $\pi$ in the systems preserving either $\mathcal{T}$ or $\mathcal{I}$. 
For three-dimensional (3D) insulators with $\mathcal T$ symmetry , the quantized $\theta$ term is related to $\mathbb{Z}_2$ invariant directly. For the centrosymmetry insulators (breaking $\cal T$ symmetry), it can be reduced to a parity-based invariant, $\mathbb Z_4=0,2$ (Note that $\mathbb Z_4=1,3$ corresponds to a semimetal with Weyl nodes), which is defined as:~\cite{turner2012quantized,watanabe2018structure,PhysRevB.98.115150}
\begin{equation}
\mathbb Z_4=\sum_{\alpha=1}^8 \sum_{n=1}^{n_{occ}}\frac{1+\xi_n(\Lambda_\alpha)}{2}~mod~4,
\end{equation}
where $\Lambda_\alpha$ are the eight $\mathcal I$-invariant momenta, $n$ is the band index, $n_{occ}$ is the total number of electrons and $\xi_n(\Lambda_\alpha)$ is the parity eigenvalue ($+1$/$-1$) of the $n$-th band at $\Lambda_\alpha$. A centrosymmetric insulator with $\mathbb Z_4=2$ can have quantized bulk TME effect ($\theta=\pi$), also called an axion insulator (AI). It has half-integer quantum Hall effect (QHE) on the surface as long as the surface states are gapped.~\cite{essin2009magnetoelectric}. Very recently, axion insulators were also proposed to host one-dimensional (1D) chiral states on the hinges, and can at the same time be viewed as the newly proposed higher-order topological insulator (HOTI).~\cite{benalcazar2017electric,langbehn2017reflection,benalcazar2017quantized,song2017d,schindler2018higher,schindler2018higher1,ezawa2018magnetic,yue2019symmetry,okuma2019topological,wang2018higher,wieder2018axion}

Although in the literature there are already a number of  material proposals for the axion insulator phase  including the heterostructure constructed by quantum anomalous Hall insulators,~\cite{xiao2018realization,mogi2017magnetic} the magnetically doped 3D TIs,~\cite{li2010dynamical,yue2019symmetry}, the axion insulator phase induced by Coulomb interaction in magnetic osmium compounds,~\cite{wan2012computational} and the layered antiferromagnetic(AFM) TI with type-IV magnetic space group(MSG),~\cite{essin2009magnetoelectric,zhang2018topological,li2018intrinsic,gong2018experimental,chowdhury2019prediction} a satisfactory stoichiometric material system with accessible single crystal is still absent, which greatly impeded
the experimental studies on the AI and topological magneto-electric effects.

In the present letter, we predict that EuIn$_2$As$_2$ is an AFM axion insulator (breaking $\mathcal T$ symmetry, but preserving $\mathcal I$ symmetry), no matter which direction the magnetic order is in. Its nontrivial topology is characterized by the parity-based invariant, $\mathbb Z_4=2$. The quantized bulk TME effect can be expected in the material with a bulk energy gap of $\sim 0.1$~eV.
As there is lack of experimental data to determine the direction of the magnetism (which {\it does} effect the surface states), we have investigated two AFM phases ({\it i.e.}, afmb and afmc phases) in the first-principles calculations. The results show that the energy difference of the two is less than 1.0 meV. 
The afmb phase of EuIn$_2$As$_2$ is described by a type-I MSG, $Cmcm$. Our calculations show the afmb phase is also a topological crystalline insulator (TCI), characterised by {\it nonzero} mirror Chern numbers (MCN) defined on the $k_y=0$ and $k_z=0$ planes. As a result, those surfaces perpendicular to one of the mirror planes are gapless. 
However, the afmc phase belongs to a type-III MSG, $P6_{3'}/m'm'c$, and no symmetry-protected gapless surface state is found for both (100) and (001) surfaces. As a consequence of a HOTI ($\mathbb Z_4=2$), a chiral mode can emerge on the hinge between those gapped surfaces, which has also been obtained in our detailed calculations. 
As EuIn$_2$As$_2$ is easy to grow and the magnetism is intrinsic and confirmed already by experiments, it's an ideal platform to study the axion insulators and HOTIs experimentally.

\begin{figure}[htbp] 
\centering\includegraphics[width=3.4in]{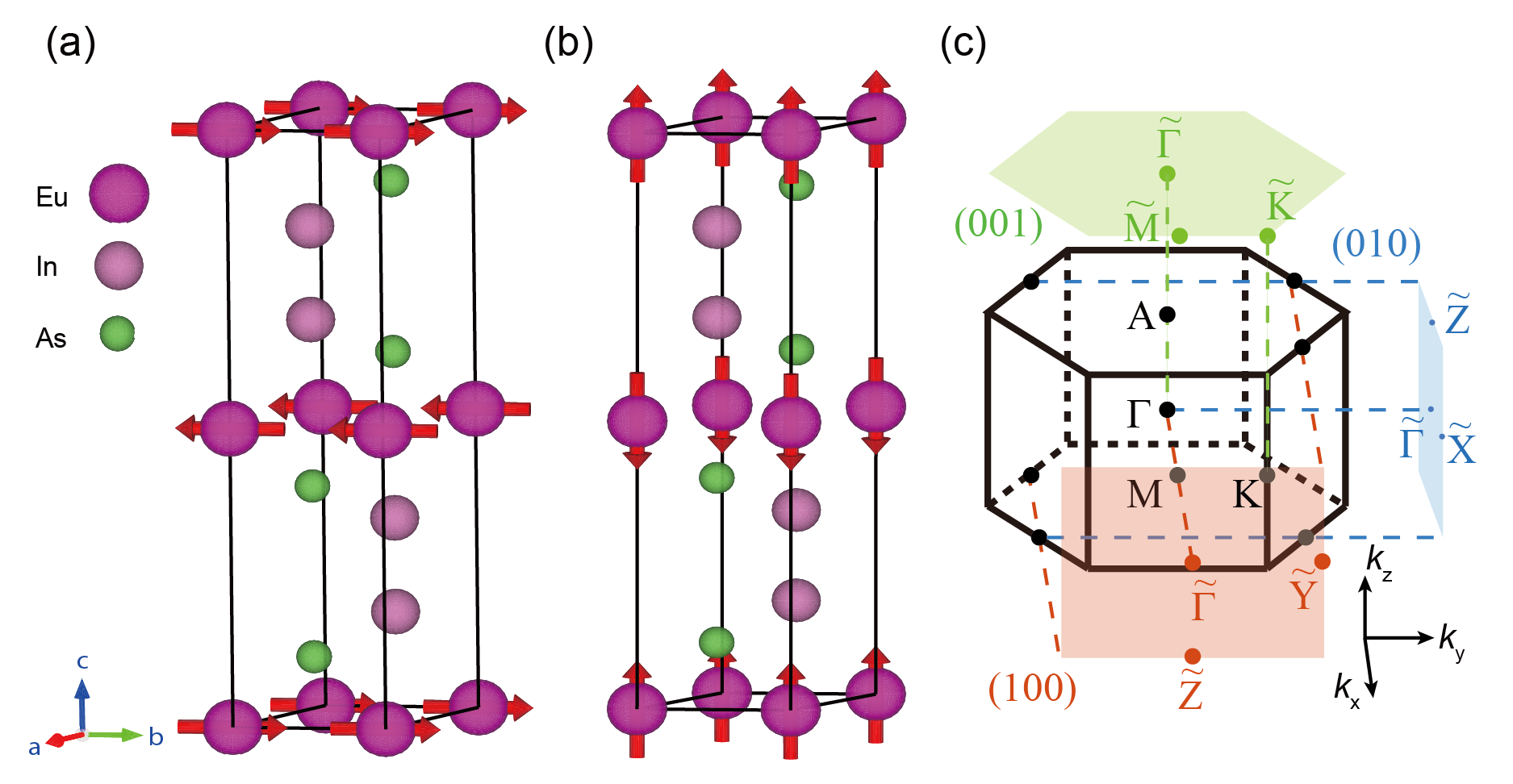} 
\caption{The Crystal and magnetic structure of EuIn$_2$As$_2$. (a) and (b) are the afmb and afmc phase, respectively.
 The crystal axes $b$ and $c$ are along $y$ and $z$ directions in Cartesian coordinate system, respectively. (c)The bulk BZ and the surface BZ projected onto (100), (010) and (001) surfaces. Here (100) ((010) or (001)) refers to the surface normal vector in terms of the Cartesian coordinate.
}\label{fig:1} 
\end{figure} 

\textit{Crystal structure and methodology. }
EuIn$_2$As$_2$ is an antiferromagnetic Zintl compound with the N\'{e}el temperature $T_N=16 K$
.~\cite{goforth2008magnetic,singh2012high} The magnetism measurements on single crystal samples show that the moments on $\textup {Eu}^{2+}$ cations are about 7.0$\mu_{B}$ indicating the $\textup {Eu}^{2+}$ ions are in
the high spin configuration of $\textup 4f^7$.
 The intra-layer exchange coupling among $\textup {Eu}^{2+}$ ions is ferromagnetic while the interlayer coupling is antiferromagnetic, leading to A-type 
 antiferromagnetic long range order in the material.
The direction of spin polarization detected by the exepriments can be along the crystallographic $a/b$-axis (afmb phase as shown in Fig.\ref{fig:1}(a)) or $c$-axis (afmc phase as shown 
in Fig.\ref{fig:1}(b)) with a small energy difference between them. 
The crystal structure of EuIn$_2$As$_2$ adopts hexagonal lattice
with the space group of P6$_3$/mmc (No. 194), as shown in Fig.\ref{fig:1}(a)(b).
It can be generated by several symmetries: $\mathcal I$, $C_{3z}$, $C_{2y}$ and $\{C_{2x/z}|00\frac{c}{2}\}$, where $(00\frac{c}{2})$ is a half of the lattice translation in the $c$ direction.
 The experimental lattice constants were used in the following calculations with $a=b=4.2055\textup\AA$
and $c=17.887\textup\AA$. The Eu atom is located at the
Wyckoff position $2a$ (0, 0, 0). Both In and As are at the $4f$ Wyckoff
position with the coordinates of (2/3, 1/3, 0.17155) and (1/3, 2/3,
0.60706), respectively. We have performed the first-principle calculations
using the Vienna ab-initio simulation package(VASP)~\cite{kresse1996VASP,kresse1993VASP} and the generalized
gradient approximation (GGA)~\cite{perdew1996generalized} with the Perdew-Burke-Ernzerhof (PBE)~\cite{blochl1994PAW,PBE1996PBE}
type exchange-correlation potential was adopted. The Brillouin zone(BZ)
sampling was performed by using k grids with a 11$\times$11$\times$3
mesh in self-consistent calculations. As the on-site Coulomb interactions among electrons on Eu-$4f$
orbitals are very strong, we have taken $U_{4f}=5.0$ eV as a parameter in the GGA+U calculations~\cite{Dudarev1998} 
to locate the occupied $4f$ orbitals at the energy determined by experiments. To calculate the topological
surface states and hinge states of antiferromagnetic EuIn$_2$As$_2$,
we have generated the maximally localized Wannier functions for $5s$
orbitals on In and $4p$ orbitals on As using Wannier90 package.~\cite{Souza2001Maximally-PRB}

\begin{figure}[htbp] 
\centering\includegraphics[width=3.4in]{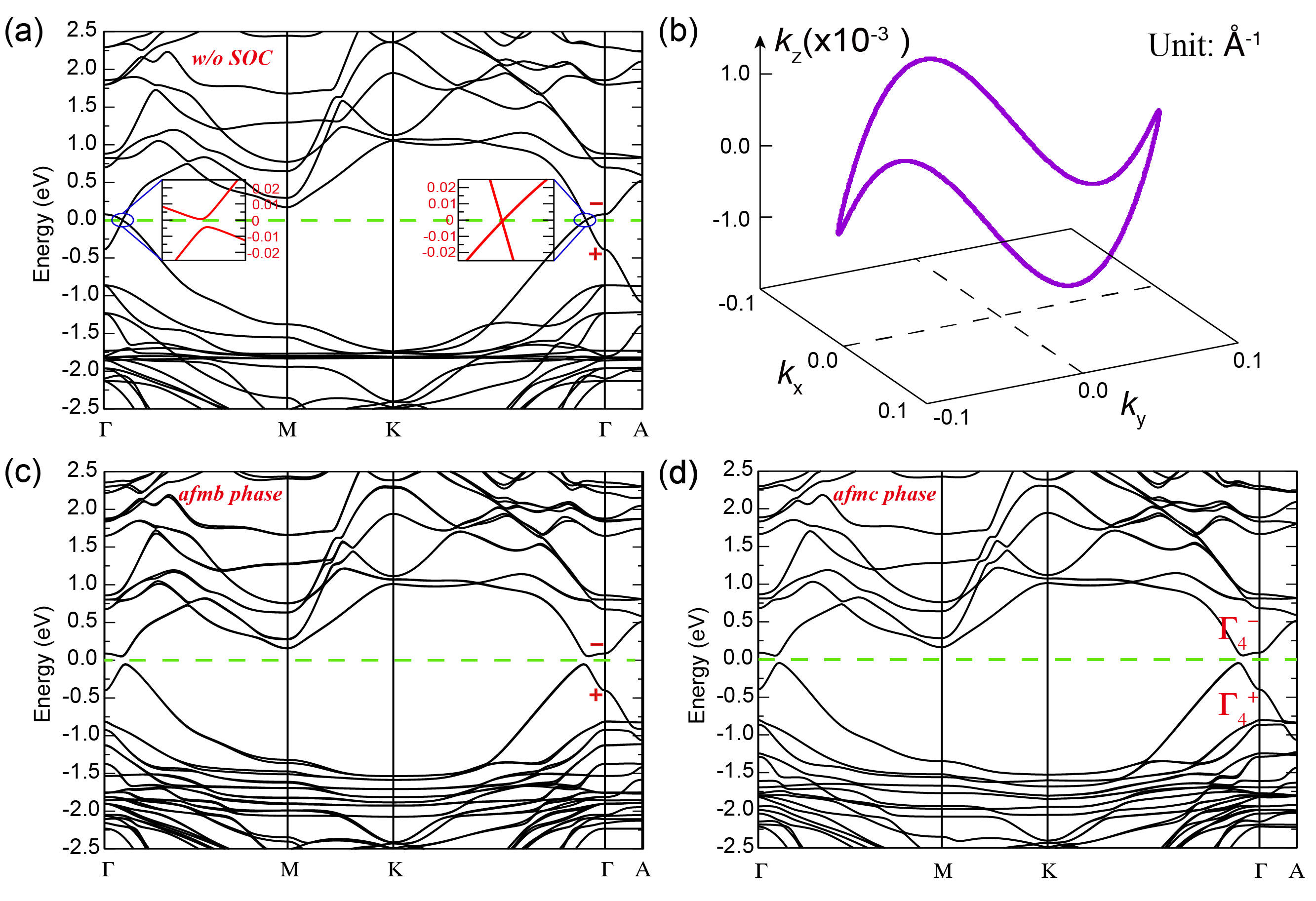} 
\caption{Electronic band structures along hingh-symmetry path in  BZ obtained by GGA+U calculations with $U_{4f}=5.0$~eV. (a)AFM band structure without SOC. The left inset shows the band crossing along $\Gamma$M is fully gapped. While the right inset shows a nodal point along $\Gamma$K path. The parities of the inverted bands have been signed. (b) Symmetry  protected nodal line near $\Gamma$ point for the spin-up channel. It is shaped by the point group $D_{3d}$ on $\Gamma$. (c) and (d) are the band structures with SOC for afmb and afmc phases, respectively. The minimal energy gap of them is about 0.1 eV. }\label{fig:2} 
\end{figure} 

\begin{table*}[t]
\caption{The number of occupied bands of odd/even parity at the eight TRIM points, without (w/o) SOC and with (w/) SOC, respectivly. The number of electrons are 146 in total.}\label{table1}
\begin{centering}
\begin{tabular}{c|p{1.4cm}p{1.4cm}p{1.4cm}p{1.4cm}p{1.4cm}p{1.4cm}p{1.4cm}p{1.4cm}}
\hline \hline
$\Lambda_\alpha$ & (0,0,0) & ($\pi$,0,0) & (0,$\pi$,0) & ($\pi$,$\pi$,0)
            & (0,0,$\pi$) & ($\pi$,0,$\pi$) & (0,$\pi$,$\pi$) & ($\pi$,$\pi$,$\pi$)
\tabularnewline
\hline
Spin-up$\ \ \ $ w/o SOC & 41/32  & 42/31  & 42/31  & 42/31  & 40/33  & 40/33 & 40/33  & 40/33  \tabularnewline
Spin-down w/o SOC & 41/32  & 42/31  & 42/31  & 42/31  & 33/40  & 33/40 & 33/40  & 33/40  \tabularnewline
\hline
w/ SOC & 82/64  & 84/62  & 84/62  & 84/62  & 73/73  & 73/73 & 73/73  & 73/73  \tabularnewline
\hline \hline
\end{tabular}
\end{centering}
\end{table*}

\textit{Nodal-line Semimetal without SOC.} We have first calculated the band structure of EuIn$_2$As$_2$ with antiferromagnetic phase in the absence of spin-orbit coupling (SOC). As the magnetic moment of the Eu atoms makes the two atoms within one unit cell not equivalent any more, the symmetry is decreased to $D_{3d}$, generated by $\mathcal I$, $C_{3z}$ and $C_{2y}$, which is the symmetry that the spin-up/spin-down bands respect. 
 The analysis of orbital character shows that the bands near Fermi energy are dominated by $5s$ orbitals of In and $4p$ orbitals of As.
 Due to the correlation effect, the fully spin-polarized $4f$ orbitals of each Eu are pushed down to $1.5\sim2.0$ eV below the Fermi level
, which is quite consistent with the position suggested by experimental measurements.~\cite{richard2013angle,gui2019new}
The band dispersion along the high-symmetry directions in Fig.\ref{fig:1}(c) is shown in Fig.\ref{fig:2}(a), 
which shows that there is a band crossing feature near $\Gamma$ point on the Fermi energy. 
Detailed calculations and further symmetry analysis shows that it is actually an anti-crossing along $\Gamma M$, while the crossing along $\Gamma K$ is protected by $C_{2y}$ symmetry, as shown in the insets of Fig.~\ref{fig:2}(a). Namely, two crossing bands belong to different $C_{2y}$ eigenvalues $\pm 1$ along the $\Gamma K$ line.

The $\mathbb Z_4$ invariant ($n_{occ} = 146/2$ for each spin channel) is computed to be 1 for the spin-up channel, which implies that there are either nodal lines or odd pairs of Weyl nodes depending on the symmetry. 
In the collinear magnetic calculations without SOC, the Hamiltonian for each spin channel is still real. There is a symmetry ($C$) of complex conjugation, $C=\cal K$, which is the same as the time reversal symmetry for spinless systems. Thus, the combined anti-unitary symmetry of $C$ and $\cal I$ satisfies $(C{\cal I})^2=1$, which forbids one of the three Pauli matrices in a two-band effective model at {\it any} $k$-point. Therefore, the nodal line is guaranteed by $\mathbb Z_4=1$. In fact, the crossing point on $\Gamma K$ is part of the nodal line, as shown in Fig.~\ref{fig:2}(b). The nodal line for the spin-down channel is obtained by a mirror reflection with respect to the $k_z=0$ plane, because the two spin channels are related to each other by the combined symmetry ${\cal T}\{C_{2z}|00\frac{c}{2}\}$.

\textit{Anixon insulator with SOC.} 
As suggested by the previous experiments ~\cite{goforth2008magnetic,singh2012high}, two meta-stable AFM phases, with the magnetic moment $m||b$-axis (afmb phase) and $m||c$-axis (afmc phase) respectively, have been investigated in our calculations including SOC. These two states are nearly degenerate from our calculations with the energy difference between the two being less than 1 meV. As the bands of 4$f$ orbitals are localized far away from the Fermi level, the low-energy band structures of the two AFM phases are very similar.
We note that SOC has two main effects: (i) it opens a band gap, which can be seen to be $\sim 0.1 eV$ in Fig.\ref{fig:2}(c) and (d); (ii) it changes the symmetry of the AFM phases, which does effect the surface states.

Due to the presence of $\mathcal I$ symmetry, one can compute the $\mathbb Z_4$ invariant , which is defined on the parity eigenvalues at the eight $\mathcal I$-invariant points $\Lambda_\alpha~(\alpha=1,2,\dots,8)$. The numbers of odd/even bands are lised in Table \ref{table1}. The obtained invariant $\mathbb Z_4=2$ suggests that EuIn$_2$As$_2$ is a axion insulator exhibiting quantized TME effect ($\theta=\pi$) without $\mathcal T$ symmetry. 
Since the change of the direction for magnetic moments won't affect the order of the bands at all the TRIM points, both of the afmb and afmc phases are AI and HOTI at the same time, which will lead to possible chiral hinge states at the hinges between two gapped surfaces.

\begin{figure}[htbp] 
\centering\includegraphics[width=3.4in]{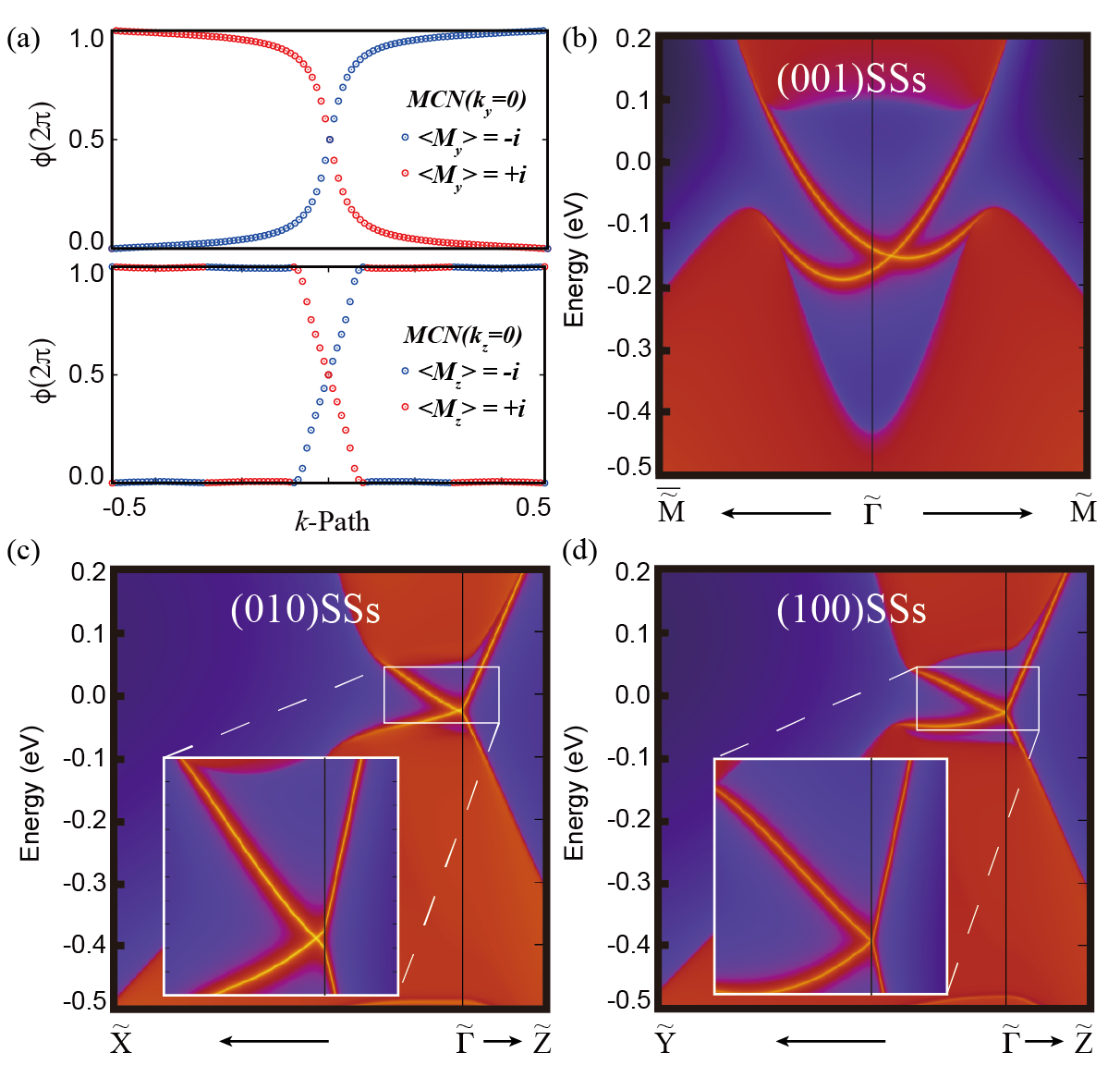} 
\caption{Numerical calculations of the mirror chern numbers and topological surface states of afmb phase. (a)Top panel: Flow chart of the average position of Wannier charge centers calculated by Wilson-loop method for the occupied bands with mirror eigenvalue $<M_y>=i(-i)$ on the mirror plane $k_y=0$, which is plotted by red(blue) circles. Bottom panel: Flow chart of the average position of Wannier charge centers calculated by Wilson-loop method for the occupied bands with mirror eigenvalue $<M_z>=i(-i)$ on the mirror plane $k_z=0$, which is plotted by red(blue) circles. 
(b)-(d) are the surface states of afmb phase along the high-symmetry lines on surface BZ of (001), (010) and (100) surface planes, respectively. (b) The Dirac cone of (001) surface is on the path of $k_y=0$ protected by $M_y$. (c) The   Dirac cone of (010) surface is on the path of $k_z=0$ protected by $\{M_z|00\frac{c}{2}\}$. (d) The Dirac cone of (100) surface is constrained on $\Gamma$ point by $M_y$ and $\{M_z|00\frac{c}{2}\}$ symmetries.}\label{fig:3} 
\end{figure} 

\textit{Topology of afmb phase. } For the afmb phase of EuIn$_2$As$_2$, the generators of the MSG include spacial inversion ($\mathcal{I}$), a two-fold rotation ($C_{2y}$) and a screw rotation ($\{C_{2z}|00\frac{c}{2}\}$).
Among the eight symmetry operations constructed from the generators, the following symmetries are important in the study of the afmb phase: $M_{y}\equiv \mathcal I C_{2y}$ and $M_{z}\equiv \mathcal I \{C_{2z}|00\frac{c}{2}\}$.
Beside the $\mathcal I$-based invatiant $\mathbb Z_4=2$, we have also calculated the mirror Chern number~\cite{hsieh2012topological} by calculationg the flow of the Wannier charge centers (WCCs) on the mirror-symmetric plane in both $+i$ and $-i$ mirror eigenvalue subspaces. Our {\it ab-initio} calculations show the $k_z=0~(k_y=0)$ plane has a nontrivial mirror Chern numbr $n_{M_z=\pm i}=\mp 1~(n_{M_y=\pm i}=\mp 1)$, while the $k_z=\pi~(k_y=\pi)$ is trivial. Namely, the afmb phase of EuIn$_2$As$_2$ is also a TCI phase with antiferromagnetic order.
The nontrivial mirror Chern numbers $n_{M_z=\pm i}=\mp 1~(n_{M_y=\pm i}=\mp 1)$ gurantee the existance of a massless Dirac cone on the $M_z$($M_y$)-preserving surface. The energy spectrum near the Dirac cone on (100), (010) and (001) surfaces have been calculated with Green's function method~\cite{sancho1985highly,WU2017}, as shown in Figure \ref{fig:3}(b)-(d). The massless Dirac cone is constrained on the path with $M_y$ for (001) surface and on the path with $\{M_z|00\frac{c}{2}\}$ for (010) surface. While for (100) surface, the Dirac cone is localized on $\Gamma$ point constrained by $M_y$ and $\{M_z|00\frac{c}{2}\}$ at the same time.The above numerical results confirmed that due to the co-existing nonzero mirror Chern number in afmb phase, all the low index surfaces like (100), (010) and (001) are gapless and the chiral hinge states can only possible on the hinges formed by two high-index surfaces which are gapped. 

\begin{figure}[htbp] 
\centering\includegraphics[width=3.4in]{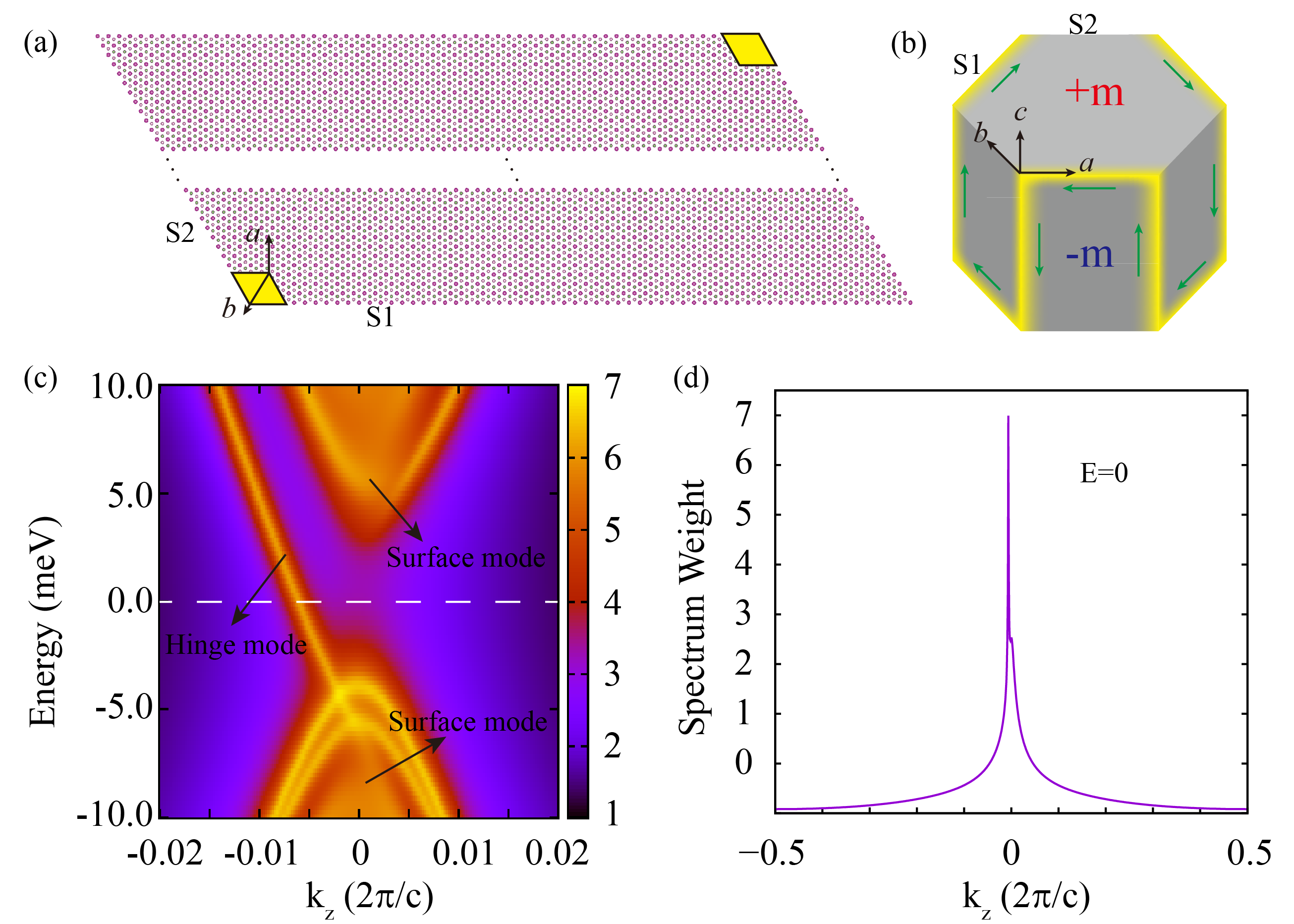} 
\caption{Chiral hinge states of the afmc phase. (a)The structural configuration used in our hinge modes calculation, which have open boundary conditions in the S1 and S2 surfaces and periodic boundary conditions in the $z$ direction. We have calculated the hinge modes with the 61-unit-cell width for S1 surface and semi-infinite width for S2 surface.  The hinges that have chiral modes are represented by yellow regions. (b)Schematic of the 1D chiral modes that localized on the hinges of hexagonally shaped HOTI with $C_{3z}$ and $\mathcal{I}$. The regions of Dirac mass term with opposite sign are separated by the yellow lines, which represent the chiral current channel. (c) Numerical calculations of the hinge states with the supercell structure in (a). (d)Spectrum strength on the Fermi energy along $k_z$ direction.  }\label{fig:4} 
\end{figure} 

\textit{Higher-order topology of afmc phase. }When the magnetic moment is along $c$-axis, the generators of the MSG include spacial inversion ($\mathcal{I}$), a three-fold rotation ($C_{3z}$), a two-fold screw rotation ($\{C_{2x}|00\frac{c}{2}\}$), and a combined anti-unitary symmetry ($\mathcal{T}C_{2y}$). 
No symmetry-protected gapless surface state is found in our calculations for both (100) and (001) surfaces.
Therefore, we can generate a hexagonal cylinder in real space (as shown in Fig.\ref{fig:4}) with all the surfaces being gapped.
But the symmetry ${\cal T} C_{2y}$ can relate two neighbor side surfaces and leads to mass terms with different sign on the neighboring surfaces of the hinges along the $c$-axis. Thus, 1D Chiral models can be expected in these hinges.
Based on the tight-binding Hamiltonian, we have simulated the hinge states using Green's function method with the structural configuration in Figure \ref{fig:4}(a).  For details of the calculational method please refer to our previous paper.~\cite{yue2019symmetry} The energy spectrum projected onto the hinge is shown in Figure \ref{fig:4}(c) and (d), which indicates that the bulk and surface states are all gapped with a minimal gap of about 10meV and a chiral mode exists on the hinge. 

The existence of the chiral hinge states can also be derived from the following kp model. The irreducible representations for the two bands near the Fermi level are $\Gamma_4^+$ and $\Gamma_4^-$ at $\Gamma$, as shown in Fig. \ref{fig:2}(d), where $\Gamma_4^\pm$ are from the character table of $D_{3d}$. Based on the invarinant theory, the low-energy effective Hamiltonian at the $\Gamma$ point can be described as:
\begin{equation}
\begin{split}
H(k)&=\epsilon(k)\mathbb I_{4\times 4}+M(k)\tau_z\sigma_0+ H_1 +H_2 \\
H_{1} &=  A(k_{x}\tau_{x}\sigma_{x}+k_{y}\tau_{x}\sigma_{y})+Bk_{z}\tau_{x}\sigma_{z} \\
H_{2} &= Ck_z\tau_{z}(\sigma_{x}k_x+\sigma_yk_y)\\
& ~~~ +D\tau_{z}[\sigma_{x}(k_x^2-k_y^2)-2\sigma_yk_xk_y] 
\end{split}
\end{equation}
where $\epsilon(k)=e_0+e_1k_z^2+e_2(k_x^2+k_y^2)$, $M(k)=M_0+M_1k_z^2+M_2(k_x^2+k_y^2)$, and A,B,C and D are real. $\tau$ and $\sigma$ represent the orbital and spin space, respectively. $H_{1(2)}$ is the first (second) order term of moment $k$. In the afmc phase, $\cal T$ is broken, while the combianed anti-unitary symmetry of ${\cal T}C_{2y}$ is preserved. Under the basis, ${\cal T}C_{2y}$, $\cal I$, $C_{3z}$ and $\{C_{2x}|00\frac{c}{2}\}$ are represented by $\tau_z\cal K$, $\tau_z\sigma_0$, $\tau_{0}e^{-i\frac{\pi}{3}\sigma_z}$, and $\tau_{0}e^{-i\frac{\pi}{2}\sigma_x}$, respectively. Note that $H_1$ virtually respects $\cal T$ symmetry ($i\sigma_y \cal K$), but $H_2$ breaks it. 
Thus, the term  $ f(k) \tau_y\sigma_0$ can be added to the $k\cdot p$ Hamiltonian of the sample, which is invariant, $R[f(k)\tau_y\sigma_0]R^{-1}=f(Rk)\tau_y\sigma_0$, under the symmetries ($R={\cal T}C_{2y}$, $C_{2x}$, $\cal I$ and $C_{3z}$):
\begin{eqnarray}
\begin{split}
f({\cal T}C_{2y}k)=f(k);&~f(C_{2x}k)=f(k) \\
f(\mathcal{I}k)=-f(k); &~f(C_{3z}k)=f(k)
\end{split}
\end{eqnarray}
Thus, to the lowest order of k, $f(k)=F(k_x^3-3k_xk_y^2)$.
By projecting the bulk Hamiltonian onto the surface,~\cite{khalaf2018symmetry} the resulting mass term changes its sign under $\mathcal{I}$ symmetry and invariant under three-fold rotation symmetry $C_{3z}$.  So for the shaped crystal structure shown in Fig. \ref{fig:4}(b), there will be a chiral current indicated by the green arrow. Therefore the HOTI phase in the present system is similar to the case in Bismuth, but it's chiral type. So the chiral hinge states are practical to be observed in STM experiments.

\textit{Conclusion. }In summary, we propose that the 3D antiferromagnetic material EuIn$_2$As$_2$ is an axion insulator with quantized TME effect in the bulk. Its nontrivial topology is characterized by the parity-based invariant $\mathbb Z_4=2$. With the easy axis along $a/b$-axis, it's also a TCI with {\it nonzero} mirror Chern numbers. Its gapless surface states have been calculated on both side and top surfaces, protected by mirror symmetries. When the easy axis is along $c$-axis, the (100) and (001) surfaces are gapped and it is also a HOTI with all the side surfaces are fully gapped. We have also calculated the chiral hinge modes in this material systems, which exist on the hinges of the crystal with hexagonal shape. 
The HOTI phase and its chiral hinge modes in the present system are protected by inversion symmetry with $\mathbb Z_4=2$.
 As EuIn$_2$As$_2$ is easy to grow and the magnetic configuration is intrinsic, it's an ideal candidate to study axion insulators and HOTIs. It also provides a platform to study the interplay between the magnetic structure and the topological feature of the band structure.

\textit{Note added}. We are also aware of the similar work~\cite{gui2019new} when finalising the present paper.

\section{acknowledgments}
Z.W. was supported by the CAS Pioneer Hundred Talents Program. H.W. acknowledges support from the Ministry of Science and Technology of China under grant numbers 2016YFA0300600 and 2018YFA0305700, the Chinese Academy of Sciences under grant number XDB28000000, the Science Challenge Project (No. TZ2016004), the K. C. Wong Education Foundation (GJTD-2018-01), Beijing Municipal Science \& Technilogy Commission (Z181100004218001) and Beijing Natural Science Foundation (Z180008). 
X.D. acknowledges financial support from
the Hong Kong Research Grants Council (Project No.
GRF16300918).

\bibliography{AI_ref}
\end{document}